# Long Distance Continuous Methane Emissions Monitoring with Dual Frequency Comb Spectroscopy: deployment and blind testing in complex emissions scenarios

**Authors**:
Sean Coburn[1,*]; Caroline B. Alden[1,2]; Robert Wright[1]; Griffith Wendland[1]; Alex Rybchuk[1]; Nicolas Seitz[1]; Ian Coddington[3]; and Gregory B. Rieker[1]

**Affiliations**:
[1]*University of Colorado Boulder, Boulder, Colorado, USA*
[2]*Cooperative Institute for Research in Environmental Sciences, Boulder, Colorado, USA*
[3]*National Institute for Standard and Technology, Boulder, Colorado, USA*
*Corresponding author: coburns@colorado.edu

**Abstract**
Continuous monitoring of oil and gas infrastructure is of interest for improving emissions and safety by enabling rapid identification and repair of emission sources, especially large sources that are responsible for the bulk of total emissions. We have previously demonstrated dual frequency comb spectroscopy (DCS) coupled with atmospheric modeling and inversion techniques (the DCS Observing System) as a viable and accurate approach for detection, attribution and quantification of methane emissions at distances of more than 1 km under controlled, steady emissions scenarios. Here, we present the results of validation testing designed to mimic the complexity of operational well pad emissions from oil and gas production, and the first field measurements at an active oil and gas facility. The validation tests are performed single-blind (the measurement and data analysis team are not given information about the emissions) at the Methane Emissions Technology Evaluation Center (METEC) test facility. They consist of a series of scenarios ranging from a single, steady-rate emission point to multiple emission points that include intermittent releases (the METEC "R2" tests). Additionally, we present field measurements at an active natural gas storage facility demonstrating that the system can remotely and autonomously monitor methane emissions in a true industrial setting. This field verification is in a configuration designed for continuous and long-term characterization of operational and fugitive emissions. These demonstrations confirm that the DCS Observing System can provide high-confidence continuous monitoring of emissions from complex, operational facilities among natural gas infrastructure.

**Introduction**
Methane is a potent greenhouse gas with a 20-year global warming potential (GWP) approximately 85 times greater than $CO_2$, and anthropogenic sources of methane contribute up to 65% of total global emissions [1]. Recent studies suggest that emissions from oil and gas production are higher than previously estimated, and demonstrate a "fat tail" distribution in which a relatively small number of large emitters contribute most of the overall methane loss to the atmosphere [2–5]. However, the variability of emissions in time is not well understood, which poses a challenge for both accurately assessing total emissions and understanding their associated safety and health risks. Emissions variability also introduces a challenge for differentiating process emissions (vents) and fugitive events (leaks), complicating efforts to find and fix large leaks efficiently. These gaps in our understanding and capabilities are due, in part, to limitations inherent to current detection technologies and approaches, which provide snapshot-in-time data, often on an infrequent basis. For example, current leak detection and repair (LDAR) work practice involves individual site visits by an operator with a handheld devices [6]. Truck-, aircraft- or drone-mounted sensors [7–13] can offer a faster alternative, but are still limited in the practicality for providing, continuous monitoring of infrastructure. Continuous monitoring would enable rapid find and fix capabilities for the largest emitters and is critical for accurate characterization of sector-wide emissions, given observed intermittency, unpredictability, and even seeming stochasticity in large emitters [14–16]. Fixed ground-based sensors hold the capability to enable continuous observations which would fill this critical information gap [17–26].

Dual frequency comb spectroscopy (DCS) in conjunction with atmospheric modeling and inversion methods (the DCS Observing System) provide accurate and continuous detection, attribution and quantification of methane emissions from oil and gas equipment [24–26], offering a promising solution to this problem. This spectrometer consists of a single, centralized DCS that samples the surrounding region via an optical transceiver that sends laser light over long, open atmospheric paths to an array of strategically positioned retroreflectors. This approach can provide continuous measurements of specific emissions sources across areas of 10 $m^2$ to 10 $km^2$ with a single instrument. The system operates autonomously, and due to its use of an active laser source, is capable of operation day and night in all temperatures and in all ground and sky cover conditions, except for dense falling snow, rain or fog. In addition, the large range of coverage enables deployments where many sites within a region and observed by a single system. These characteristics together make for efficient, continuous monitoring and rapid identification of emissions in operational scenarios and enable characterization of the time variability of emissions in fundamental studies.

Previous testing of the DCS Observing System at the Methane Emissions Technology Evaluation Test Center (METEC) demonstrated high-accuracy steady emission detection and characterization [25]. In a series of 18 single steady-leak tests (emission rates ranged from 0 to 10.7 g $min^{-1}$ [0-34.7 scfh]), the DCS Observing System successfully identified all emissions and quantified emission rates to within 27% on average. However, limitations of that study for applicability of the system to "real-world" emission scenarios included the use of: 1) steady emission rates, whereas intermittent or episodic emissions are observed in oil and gas operational environments (e.g., liquid unloading events [27]), and 2) single point source emissions at a time, whereas at real oil and gas facilities many components have the potential to emit simultaneously. Here we validate the DCS Observing System's performance under time-varying and multi-point source emissions scenarios and demonstrate autonomous emission retrievals in a true industrial setting.

Validation was accomplished through a set of blind tests of complex "operational" emission scenarios. The validation tests were again conducted at the METEC facility using their "R2" protocol, which included emissions that vary with time and simultaneous emissions from multiple source locations. Additionally, the site and equipment being monitored during the tests were configured to more closely match modern upstream oil and natural gas facility layouts, in terms of the size of the monitored area, and the count and spacing of emission sources. We also present data collected at an operational natural gas storage facility during a transient emission event. These measurements represent the first set of results from a long-term study of this facility and demonstrate the capability and utility of the DCS Observing System in providing time-resolved emission measurements over extended periods of time.

**Methods**
Frequency combs are laser sources that emit a large number of discrete, evenly spaced, wavelengths (comb teeth), which can be used to measure the concentration of gases in the atmosphere. This is done by transmitting comb light across an open path and detecting the wavelength-dependent absorption fingerprints of the molecules present in the path. We use dual frequency comb spectroscopy (DCS), a technique in which two carefully prepared frequency combs are interfered on a single photodiode allowing for broadband, high-resolution spectroscopy with a simple detection scheme [28,29]. In the implementation here, the frequency combs are mode-locked erbium fiber lasers that are spectrally broadened and then optically filtered to 40 nm [~4.5 THz ] to match the methane absorption fingerprint at 1.65 µm. The combs provide > 22 000 distinct spectral elements in this band at a very close 1.8 pm spacing. This enables trace gas spectroscopy over long pathlengths with high precision and stability [30,31]. The DCS used for this work has been engineered to support stable operation in field deployment [24,32].

To observe the surrounding environment, DCS light is fiber coupled to an optical telescope transceiver that launches the laser light over long atmospheric paths (typically several hundred meters to multiple kilometers) to a retroreflector, which returns the laser light back to the transceiver, where it is coupled onto a 150 MHz InGaAs photodetector. The transceiver is mounted on a gimbal that cycles the beam over multiple paths across a region, enabling the measurement of path-integrated methane concentrations over multiple-square-kilometer regions. The DCS observing system is configured such that each beam path is queried autonomously during active measurement periods and trace concentrations are derived for each measurement along a particular beam path. A more detailed description of the fielded DCS observing system and typical site configurations can be found in [24,25].

To quantify emission sources the path-integrated methane concentration measurements are incorporated into an atmospheric model and inversion algorithm allowing determination of emission location and rate. The inversion model also incorporates local wind measurements collected by a 3D sonic anemometer as well as geospatial information including retroflector locations, the DCS location, and locations of potential sources. A gaussian plume model is used to create source-receptor relationships which are then fit to the measured methane concentrations using a least-squares-based inversion approach providing the emission location and rate. Further detail can be found in [26]. Note that the number of individual beam paths and total number of samples utilized by the inversion is determined by the system configuration and goals of the measurements. For the measurements presented here, the number of used beam paths varied from 2 - 10 and total sample times are on the order of 2 – 4 hours.

**Blind validation testing under "operational" conditions**

The Methane Emissions Technology Evaluation Center (METEC), located in Fort Collins, CO, is a facility which was specifically designed and built to test and validate methane emissions detecting technologies. The outdoor facility is comprised of decommissioned oil and natural gas infrastructure which has been modified to enable controlled releases of gas mixtures from various points throughout the equipment (https://energy.colostate.edu/metec/). Additionally, at the time of testing the facility had recently upgraded their testing profile for more complex emissions meant to more accurately simulate realistic oil and gas operational conditions. This evaluation followed the "R2" testing procedure and consisted of 15 single-blind emission scenarios administered by METEC (each lasting ~4 hours resulting in 2 tests/day) in three "difficulty" categories:

    **Difficulty A:** Single steady emission (6 tests)
    **Difficulty B:** Multiple steady emissions (6 tests)
    **Difficulty C:** Multiple emissions that may be steady or intermittent, i.e. the rate is variable in time (3 tests)

All emission points for these tests originate from a 50 m × 60 m mock production pad containing three equipment batteries (a battery of 5 wellheads, a battery of 4 separators, and a battery of 3 tanks – where "battery" is defined here as an equipment grouping of the same type). Figure 1 shows an overhead view of the emission test site relative to the DCS location > 1 km away at the Colorado State University Foothills campus (panel a) and a more detailed view of the METEC site contained within the inset. The METEC facility is located ~5 km west-northwest of Fort Collins, CO and is not substantially impacted by any local sources of methane; however, background methane concentrations did fluctuate depending on wind conditions (although not necessarily correlated). The total range of background concentrations experienced was typically on the order of 100 – 200 ppb throughout the testing time period (~8 hrs / day).

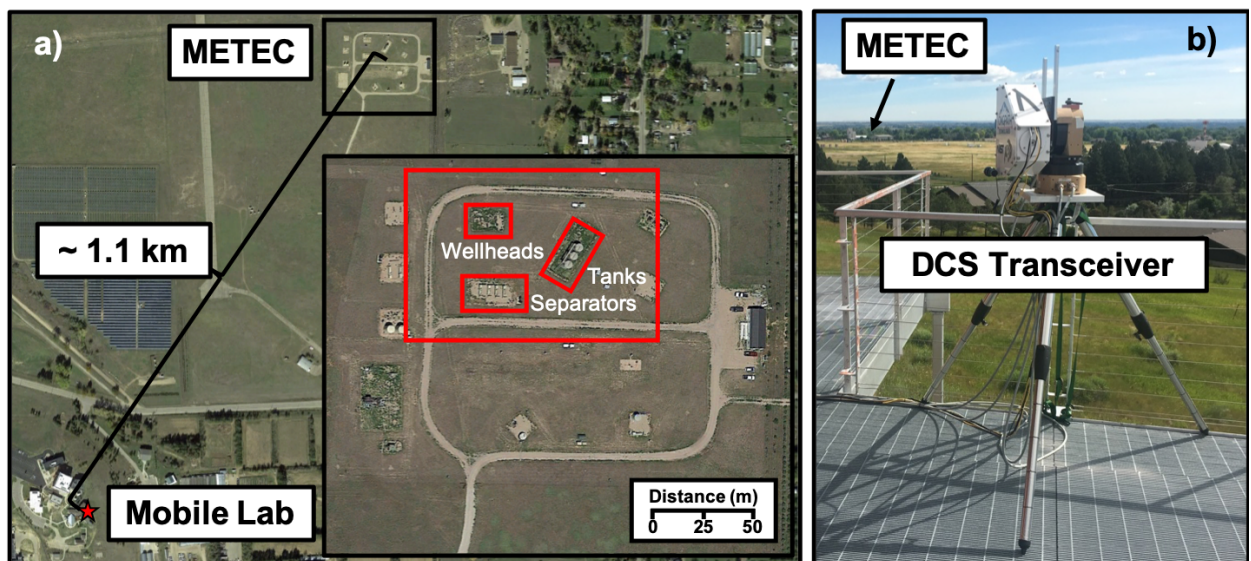

**Figure 1:** Overview of the METEC test site and the DCS sampling location (panel a) and a photo of the DCS transceiver (panel b). Panel a) the red star indicates the position of the DCS at the foothills campus of Colorado State University, which is ~1.1 km away from the METEC site. The inset includes a closer view of METEC, with the red squares highlighting the different potential source regions: 1) a battery of 5 wellheads; 2) a battery of 4 separators; and 3) a battery of 3 tanks. Panel b) contains a picture of the DCS transceiver with the METEC site noted in the background. The larger red square in the inset corresponds with the region shown in figure S1. panel a.

In the results presented here, we distinguish between the values measured by the DCS system in single-blind testing as "measured" and the values controlled by METEC as "true". For each test, we describe our success in emissions characterization in terms of detection, attribution and quantification. First, we report our success in estimating the presence and absence of emissions at each battery (detection). Second, we report our success in attributing detected emissions to the battery level (and not, for example, to individual pieces of equipment). For inversion results where > 1 emission point is detected at a given battery, the rates are summed and the average location is reported (noted in figure 2 with the cross symbol). To aid in comparison of measured with true results, we treat true results in the same way (i.e., the locations for multiple emissions from a single battery are summed and noted in figure 2 with an asterisk symbol). Finally, we report our success in estimating the emission rate for each battery (quantification), following the steps above in cases of multiple emissions. For intermittent emission profiles, the true average emission rate is calculated following Eq. 1.

$$Rate_{avg} = Rate_{cal} * (\frac{t_{on}}{t_{total}}) \qquad (Eq.\ 1)$$

Where $Rate_{cal}$ is the flow rate as measured by METEC operators during a calibration period conducted prior to each test, $t_{on}$ is the total time the emission is active, and $t_{total}$ is the total test duration.

Figure 2 is a graphical table overview of the testing results with respect to detection and attribution. In all, the system successfully detects 22 of the 25 individual emission events (dark green circles) from a distance > 1 km with 3 missed emissions (orange diamonds) and 6 false positives (red triangles) - all of which were below 1.65 g min$^{-1}$ (5 standard cubic feet per hour [scfh]). In this figure, emissions are grouped by battery along the y-axis and in increasing emission strength (from any battery for each test) along the x-axis. Additionally, the black squares indicate a test/battery combination that did not have any emissions and was correctly identified as such. A more detailed overview of the results from one of the tests (Test ID #10 from figure 2) is presented in the supplementary information.

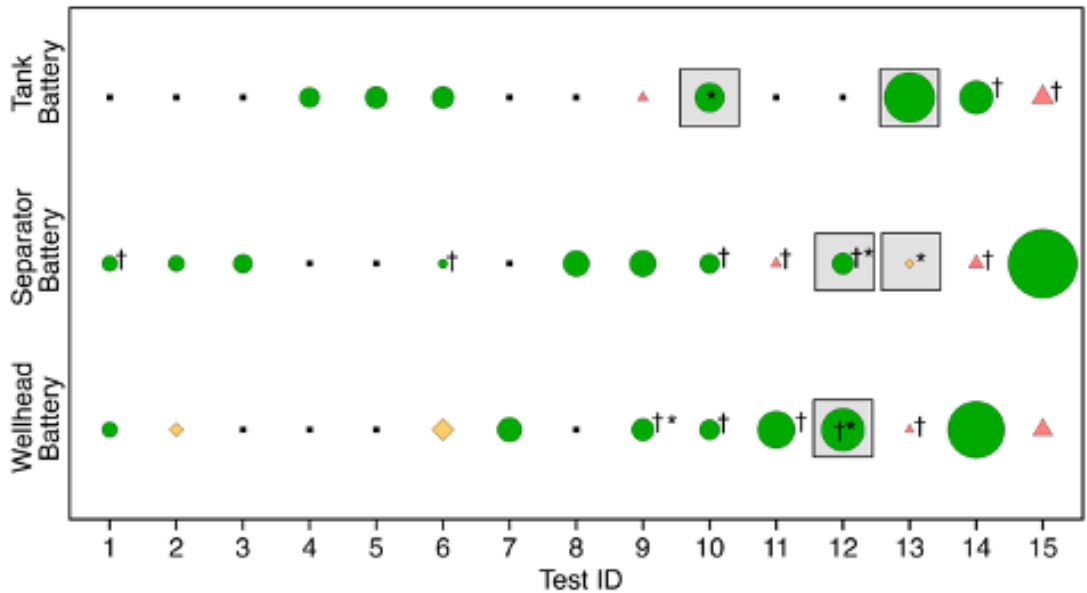

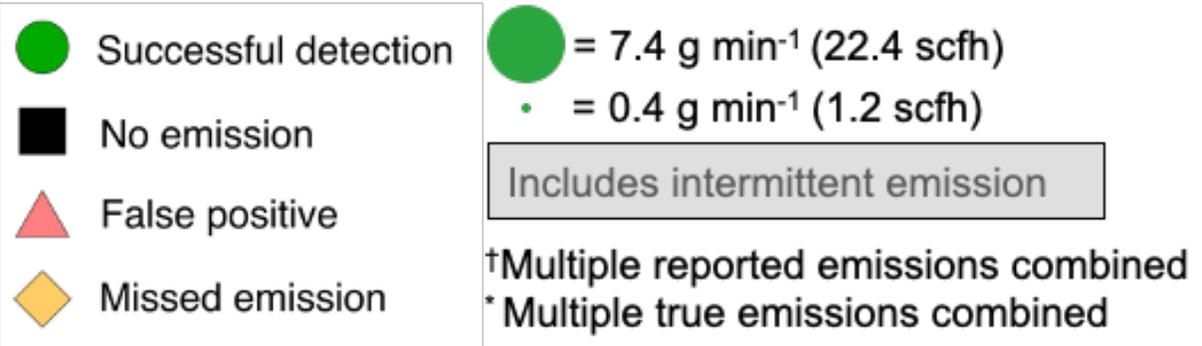

**Figure 2:** Overview of the METEC testing results. Each test is represented by a column in the figure and differentiated by a "Test ID" (ordered by the emission rate of the largest emission). Each row represents emissions at the battery level for a particular test. The shading and shape of each marker represents different outcomes where: 1) dark green circles represent a successfully detected emission point(s); 2) black squares indicate successful exclusion of a non-emitting source; 3) light red triangles indicate a reported emission when none was present (false positive); and 4) light orange diamonds indicate an undetected true emission point. The size of the marker corresponds to the size of the true emission rate for each test; except for the false positive scenarios where the size corresponds to the DCS measured rate. Additionally, tests that included intermittent emission profiles are distinguished with a gray box around the marker.

Figure 3 shows a comparison between the measured and true emission rates (quantification), grouped at the battery level, for all detected emissions. Accurate quantification of emissions to within 1.65 g min$^{-1}$ (5 scfh) is achieved in 82% of tests, and 91% of the emissions are accurately quantified to within 2.3 g min$^{-1}$ (10 scfh). Of the four tests that do not achieve 1.65 g min$^{-1}$ quantification, three include intermittent emission profiles. It should also be noted that the largest of the tested emission profiles reaches just over 7 g min$^{-1}$, which is far lower than the individual emission rates that are estimated to account for the majority of total emissions based on past field studies [3]. To understand why the intermittent emission tests are not quantified as well as the other tests, we perform a series of synthetic tests (see supplemental information), and find that monitoring for longer periods of time (~10-12 hours compared to the 4-hour duration of each test here) would likely have resulted in accurate quantification of these emissions. The need for longer monitoring times to accurately quantify intermittent emissions arises because the duration of some intermittent bursts tested by METEC was below our system averaging time (i.e., emissions only occurring for time periods on the order of seconds). It is worth noting that some intermittent emissions described in the literature are longer in duration (e.g., on the order of minutes to hours) and would not pose this need [27,33]. Even with the potential offset between the timing of intermittent emissions and the system samples, the ability to provide some level of quantification is retained due to the continuous monitoring capability of the system.

Figure 4 summarizes differences in the measured versus true emission locations for the 22 detected battery-level emission events – this is calculated as the two-dimensional difference (in the ground plane or footprint), excluding heights. 60% of the emissions are attributed to within 4 m of the true location, which roughly corresponds to attribution of the emission point to one half of the equipment battery. All emissions locations are correctly estimated to within 7 m of the true emission point (equipment battery level). The level of granularity in localization from this system is correlated with the number and placement of individual beam paths; in this scenario, the level of attribution is achieved with only a single set of beam paths (1 – 3 typically) parsing each battery. As shown in [25], higher levels of detail can be attained for attribution through the use of more beams. Additionally, if a large enough number of unique wind vectors are sampled, attribution between multiple potential sources within a particular beam pair becomes possible due to more robust model fits in the inversion algorithm. The number of unique wind vectors can, but does not necessarily, increase with increasing measurement time on a single beam pair.

The single-blind tests presented here highlight the viability of the dual comb system as a robust emission monitoring method for real-world emission scenarios. The emission profiles were specifically designed by METEC personnel to be representative of those observed at operational facilities. These results demonstrate several key strengths and limitations of the dual comb system. For example, whereas extremely precise detection, attribution, and quantification of very small emissions (< 2 g min$^{-1}$) can be achieved under operational conditions from a long distance, a limitation of the system is that accurate quantification of intermittent emissions requires longer duration of monitoring. As we describe in the following section, real-world monitoring with the DCS system can leverage monitoring durations that are continuous for days to weeks or longer, and therefore reduces issues of quantification under short-duration tests.

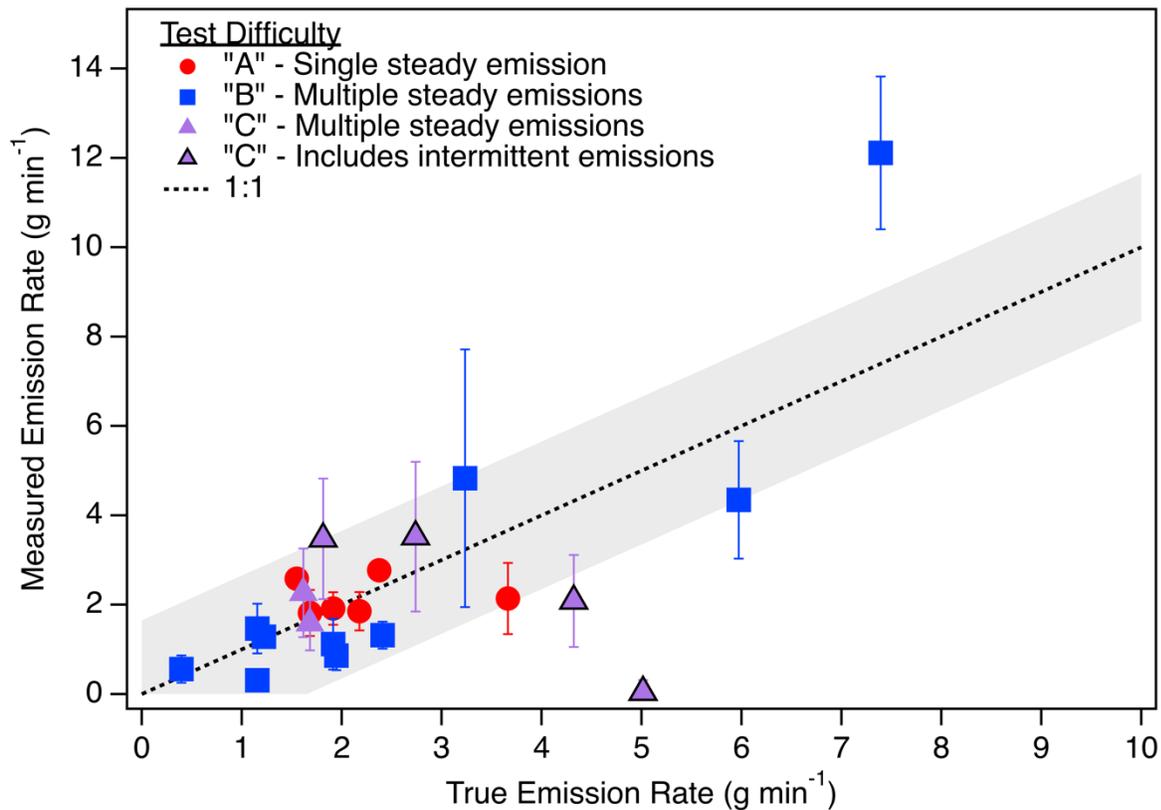

**Figure 3:** Scatter plot of the measured emission rate vs the true emission rate at the battery level for each detected emission. Data points are grouped by test difficulty (see main text): A – red circles; B – blue squares; and C – purple triangles. Further, the C level emissions which included intermittent emission profiles are denoted with a black outline around the marker. The shaded region around the 1:1 line shows +/- 1.65 g min$^{-1}$ (5 scfh).

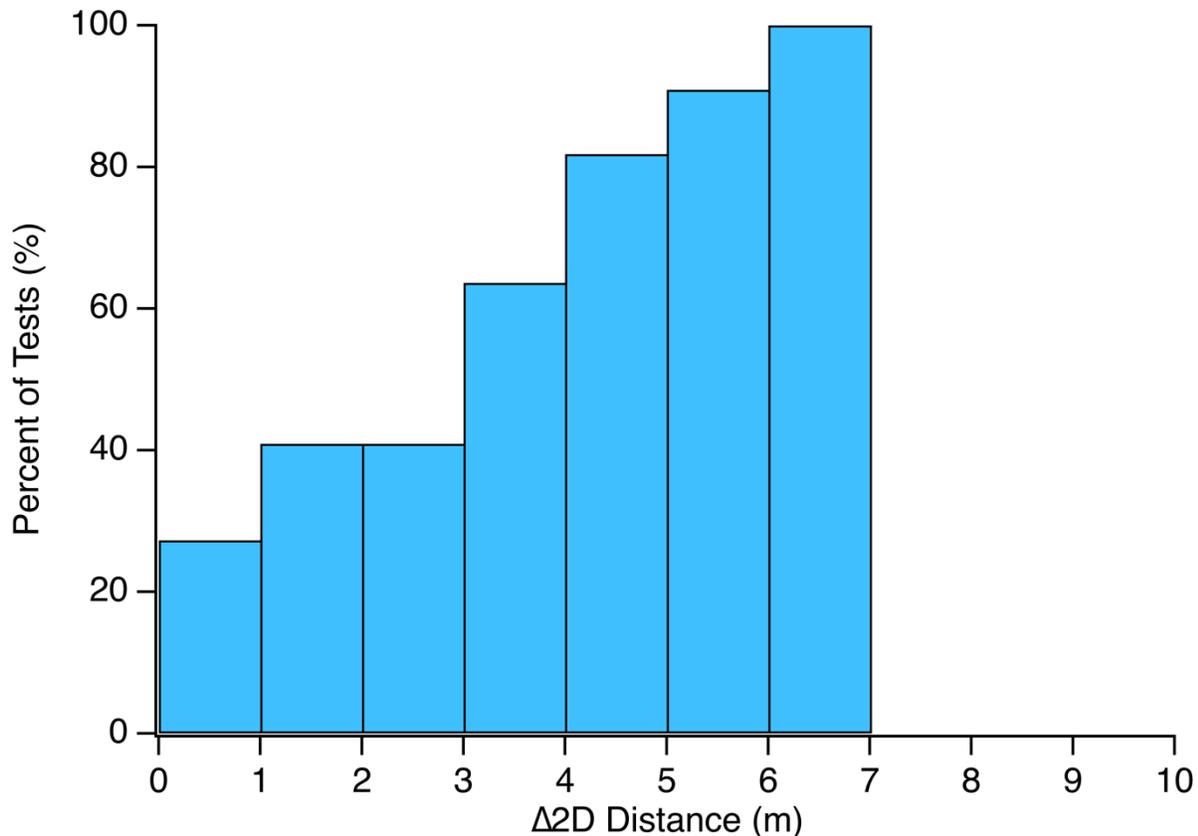

**Figure 4:** Cumulative histogram of the distance offset between the reported and true emission locations. For each distance (x-axis), the y-axis denotes the cumulative percentage of tests which were at or below each distance bin. This demonstrates that 100% of the tests were located to within 7 m of the true emission locations.

**Natural Gas Storage Site Deployment**
Currently, four DCS systems are deployed with industrial partners in the production and storage sectors. All deployments are planned to be "long-term" (6-18 months) in order to observe daily, monthly, and seasonal variability in emissions; future publications will cover in detail the deployments, measurements, and results. Here, we focus on the demonstration of the DCS Observing System for one of the natural gas storage sites as an example of how this system can be employed in a real industrial site as well as an example of the system's ability to operate autonomously and handle large changes in background methane level common at oil and gas sites.

The natural gas storage site is a moderate-size facility with several different potential sources of emissions (figure 5). Additional details about the storage site are omitted to maintain anonymity of the partner. The DCS measurements are configured to provide monitoring coverage for portion of the facility containing injection/withdrawal wells and a large natural gas-handling platform. Further details regarding the site, measurements, and study outcomes are included as part of an upcoming publication [Alden et al., in review].

The DCS is housed in a trailer located on the site with the gas-handling platform (GHP). This location is chosen to meet the goal of isolating and understanding emissions from the GHP and the two co-located banks of injection/withdrawal wells from the rest of the facility. A total of 8 beam paths are configured around the GHP - four paths per side (east/west) (figure 5). Three additional beam paths (not shown in figure 5) monitor three satellite wells that are within 800 m of the DCS system. Sampling times for each of the beam paths vary between 120-180 s depending on conditions and system performance (e.g., precipitation, condensation, etc.). The optical transceiver for the system is placed on a short (1.2 m) platform located on top of the trailer. The transceiver gimbaling system is a commercially available astronomical telescope mount. The system runs autonomously with periodic (~once per day), brief remote check-ins from Boulder, CO.

Figure 5 shows a series of DCS system measurements over a ~19 hr time frame that spans a period of variable emissions. Significant enhancements of $CH_4$ along beam paths downwind of the GHP are consistent under multiple changing wind directions (for example the 180º shift that occurs at ~15:30). The time-resolved emission rate is retrieved through the inversion process utilizing 3-hr measurement periods (chosen to match the time frame of the DCS Observing System validation tests at METEC). The system is capable of realizing other (higher or lower) temporal resolutions by incorporating more or fewer individual measurement samples in the inversion, generally trading higher uncertainty for higher time resolution, as discussed later. Additionally, it should be noted that the relationship between concentration enhancements and the resulting emission rate is non-linear due to atmospheric dispersion and advection of the plumes; i.e., similar concentration enhancement values can be attributed to different emission rates depending on atmospheric stability and wind characteristics. The inversion algorithm accounts for these meteorological characteristics when determining the emission rate.

The wind conditions and beam pattern are such that full characterization of the facility area is achieved through a majority of the day. During most 3-hour periods, emissions remain relatively steady at approximately 10 kg hr$^{-1}$. Between 09:00 - 12:00, the emission rate increases to 20 - 40 kg hr$^{-1}$ and remains higher for ~9 hrs, until returning to a lower rate between 15:00 – 18:00. Several individual, highly elevated concentration measurements reaching up to 6 – 8 ppm (path-averaged) are noted through the measurement period. These measurements can be caused by rapid, short increases in the emission rate, or by variations in atmospheric mixing. A rapid increase in emissions is most likely the cause of three of the elevated measurements given the stability of the winds during and around the time of these particular measurements. Wind speeds are low and variable around the time of the fourth elevated concentration, which may lead to variability in localized methane mixing/pooling. If these effects are not captured by the gas transport model in the inversion, there is more uncertainty in whether rapid changes in the concentration during these types of wind conditions are due to emissions changes or variability in atmospheric mixing. Thus, due to variability in emissions, atmospheric mixing and meteorological conditions, using a limited number of observations collected during a short period in the inversion may not accurately reflect the actual time average of emissions. Therefore, continuous observations over extended time frames, such as those shown here, offer the capability of determining a more representative average emission rate for a particular monitoring time period.

The timeseries of atmospheric observations in figure 5 shows that the background methane concentration can vary by up to several parts per million (ppm) throughout the day. A highly variable background signal was also observed in METEC testing. These fluctuations are likely due to atmospheric transport of near-field and far-field emission sources that are outside the monitoring domain. Rapid background changes can present issues for monitoring approaches that have lower temporal resolution or do not account for this type of rapid background fluctuation.

Attribution of emission sources to different locations/equipment (the GHP, wells and tanks, figure 5 (a)) varies throughout this sampling period with the largest emissions (between 10:30 and 13:30) coming from the southeast side of the GHP. However, further attribution of emissions during this particular time period to a single bank of equipment is hindered by the layout and monitoring conditions. Alden et al., (in review), presents a more thorough analysis of the long-term measurements and emissions from this deployment.

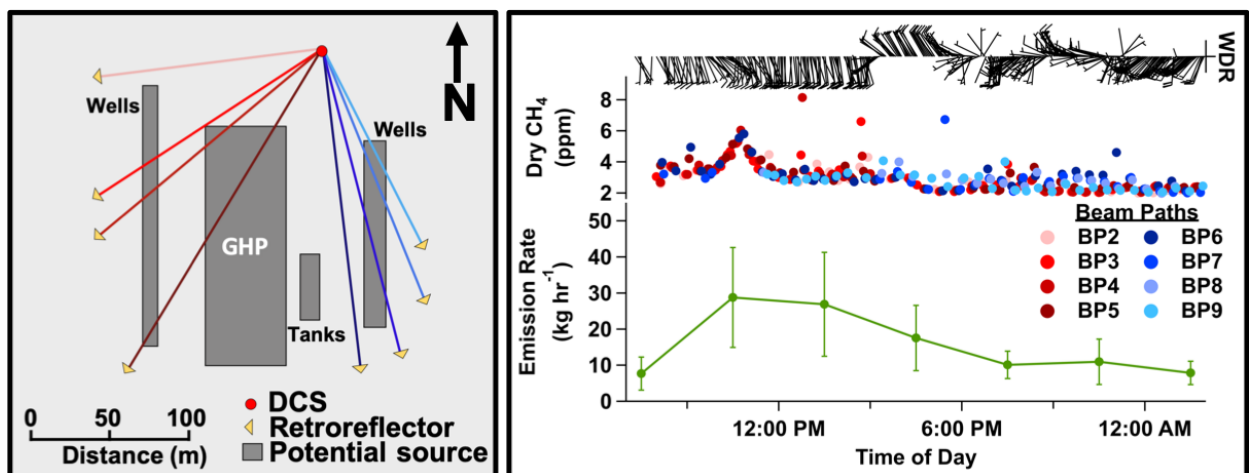

**Figure 5:** Data sample from the DCS system deployed at an active natural gas storage facility. Panel (a) shows a schematic of the site configuration with the beam paths overlaid – the beam path colors correspond to their respective traces on the time series plot. Panel (b) shows the methane mole fraction measurements for individual beam paths, the calculated emission rate (error bars represent the standard deviation of the results from the bootstrapped inversion method as per [26]), and the wind barbs (upper axis, arrows point towards the direction from which the wind was blowing and the number of barbs indicate wind speed).

**Conclusions**
We present the successful implementation and validation of a dual-frequency comb spectrometer coupled to an atmospheric inversion system (DCS Observing System) for the monitoring of methane emissions under real oil and gas emissions scenarios. We validate the DCS Observing System in a series of blinded, controlled emissions tests at the METEC test facility designed to represent operational emissions from natural gas systems. We further demonstrate successful implementation of the system at an active natural gas storage facility, where the system provides continuous emissions monitoring. The results of the two rounds of controlled testing at the METEC facility (this study and [25]) and experience gained from operating the DCS system at active natural gas handling facilities collectively validate this method and technique as a viable option for monitoring and characterization of methane emissions at temporal and spatial scales which are under-represented in currently utilized methods.

The single-blind testing conducted at the METEC facility affords the opportunity to quantitatively assess the system in a controlled setting in which complex and realistic emission scenarios are simulated. The results show that the DCS system is capable of characterizing a variety of complex emissions – including distinguishing simultaneous emissions originating from different pieces of equipment in close proximity and in some cases identifying when an emission is intermittent. Throughout the testing, 6 false positives are reported and 3 emissions go undetected – in all of these cases the emission rates are below 1.65 g min$^{-1}$ (5 scfh). This finding suggests a detection threshold for the system of roughly 1.65 g min$^{-1}$ from a distance of 1 km under complex monitoring conditions (i.e., multiple intermittent potential sources both inside and outside the immediate monitoring region). This threshold is very small in comparison to the distribution of emissions rates observed in past studies of the oil and gas production sector. For example, a previous device-level study around active oil and gas facilities showed that 90% of total emissions come from devices with emissions > 42 g min$^{-1}$ (134 scfh) (e.g. [3]).

For the single-blind tests, all emissions are correctly attributed to within 7 m of the true emission location, which enables differentiation between the different batteries of equipment and often between different components of a single battery with the laser beam configuration used here. Of the reported battery-level emission rates, 84% are quantified to within 1.65 g min$^{-1}$ (5 scfh), with the largest discrepancy (5.3 g min$^{-1}$ [16 scfh]) occurring during a test of exclusively intermittent emission profiles (Test #13, further discussed in Appendix A). Further testing suggests that for emissions exhibiting an intermittent profile, the time required to achieve accurate quantification can be greater than that for steady emissions.

This first demonstration of a DCS Observing System deployed at an active natural gas facility is intended to show the ability to provide long-term continuous monitoring of methane emissions and variability with the overarching goal of improving our understanding of process-level emissions, their variability through time, and how the system can be used as a methane emission mitigation tool. Even with the limited data presented here, several key features are clearly demonstrated from the concentration and emission time series; 1) in areas with dense natural gas operations the background level of methane can vary significantly (several ppm in these measurements) and rapidly (variations shown in this data are >2 ppm hr$^{-1}$ at some points); and 2) emission rates can also change dramatically (factor of 3 in this data) and on short time scales (hours) (figure 5). These events are not necessarily correlated (i.e., changes in background concentrations do not necessarily indicate the presence of a local emission source) and monitoring systems must be capable of distinguishing between them.

The results shown here demonstrate the system's ability to fill a critical spatial and temporal monitoring gap in methane observation technology. The DCS system described, validated, and field-deployed here provides continuous information on a regional scale. Full characterization (i.e., identifying emitting batteries and quantifying emissions rates) of multiple sources across a region (or facility) over extended time periods will aid in enhancing our understanding of methane sources from this important sector of global methane emissions.

**Funding**. Advanced Research Projects Agency-Energy, Department of Energy (award DE-AR0000539), the Department of Energy (award DE-FE0029168), and the National Institute of Standards and Technology.

**Acknowledgment**. The authors would like to thank research leaders and personnel at the Methane Emissions Technology Evaluation Center (METEC) as well as the Cooperative Institute for Research in the Atmosphere (CIRA) for aiding with the logistics of testing the system presented here.

**Data availability statement**: The data that supports the findings of this study are available upon request from the authors.

**Disclosures**: S. Coburn, C. Alden, R. Wright, and G. Rieker disclose financial interest in the company LongPath Technologies Inc., a commercial entity that employs similar trace gas sensing and inversion techniques to those used in this study.

***Supplementary Information*** to

*Long Distance Continuous Methane Emissions Monitoring with Dual Frequency Comb Spectroscopy: deployment and blind testing in complex emissions scenarios*

**Authors**:
Sean Coburn[1,*]; Caroline B. Alden[1,2]; Robert Wright[1]; Griffith Wendland[1]; Alex Rybchuk[1]; Nicolas Seitz[1]; Ian Coddington[3]; and Gregory B. Rieker[1]

**Affiliations**:
[1]University of Colorado Boulder, Boulder, Colorado, USA
[2]Cooperative Institute for Research in Environmental Sciences, Boulder, Colorado, USA
[3]National Institute for Standard and Technology, Boulder, Colorado, USA
*Corresponding author: coburns@colorado.edu

**Test ID #10 overview**
Test #10 presented one of the more complex emission profiles, with one steady emission source on each of the three batteries and one additional intermittent source on the tank battery (total of 4 distinct emission points and rates, "C" difficulty level as defined previously). Figure S1 contains a summary of Test #10 data and results. Panel (a) is a close-in view of the METEC site with the sampled beam paths overlaid; panel (b) shows the locations and nature (intermittent vs steady) of both the true and measured emissions; and panel (c) contains a time series of the methane concentration measurements along with the wind measurements shown along the top axis as wind barbs. During Test #10, a total of 7 different beam paths were sampled – a larger number than would normally be necessary due to the addition of two extra beam paths further outside the domain (beam paths 1 and 7) as well as one extra between the separator and tank batteries (beam path 5). The additional paths were added to ensure coverage for this series of tests. However, further testing with the collected data revealed that removing these extra beam paths during the inversion process did not significantly impact the system's ability to characterize these emissions.

Two important features are apparent in the methane concentration time series in figure S1 panel (c). First, the site-wide methane background varies by approximately 80 ppb during the four-hour test, as indicated by the common time variation among all 7 beam paths. Given the inherent stability of the DCS technique [1] as well as a lack of correlation between the variations and any environmental parameters that could affect the DCS, the background fluctuations are assumed to be from an off-site but local source. Second, the differences between upwind and downwind beams, which form the basis for leak detection and quantification, are between 0-20 ppb. Thus the signal from local methane leaks of the small sizes tested here are much smaller than the normal variation of the background methane concentration across the site.

Despite these challenges, the DCS system is able to accurately detect, attribute, and quantify the emissions. The true emissions during this test were as follows: wellhead battery – one steady emission at a rate of 1.7 g min$^{-1}$ (5.1 scfh); separator battery – one steady emission at a rate of 1.6 g min$^{-1}$ (4.9 scfh); tank battery – one steady emission at a rate of 1.9 g min$^{-1}$ (5.9 scfh) and one intermittent emission with an average rate of 0.8 g min$^{-1}$ (2.4 scfh). The average emission rates for all three batteries where accurately quantified to within < 1 g min$^{-1}$ with measured rates of 1.6 g min$^{-1}$ (4.8 scfh), 2.3 g min$^{-1}$ (6.9 scfh), and 3.5 g min$^{-1}$ (10.7 scfh) for the wellhead, separator, and tank batteries, respectively. Additionally, the reported emissions for each of the batteries were localized to either the correct or neighboring piece of equipment. As noted in figure S1, the wellhead and tank battery emissions were reported as intermittent and the separator emission as steady; whereas, only the tank battery contained a true intermittent emission. As described in Appendix A, accurate quantification of intermittent emissions appears to require a longer monitoring period as compared with steady emissions.

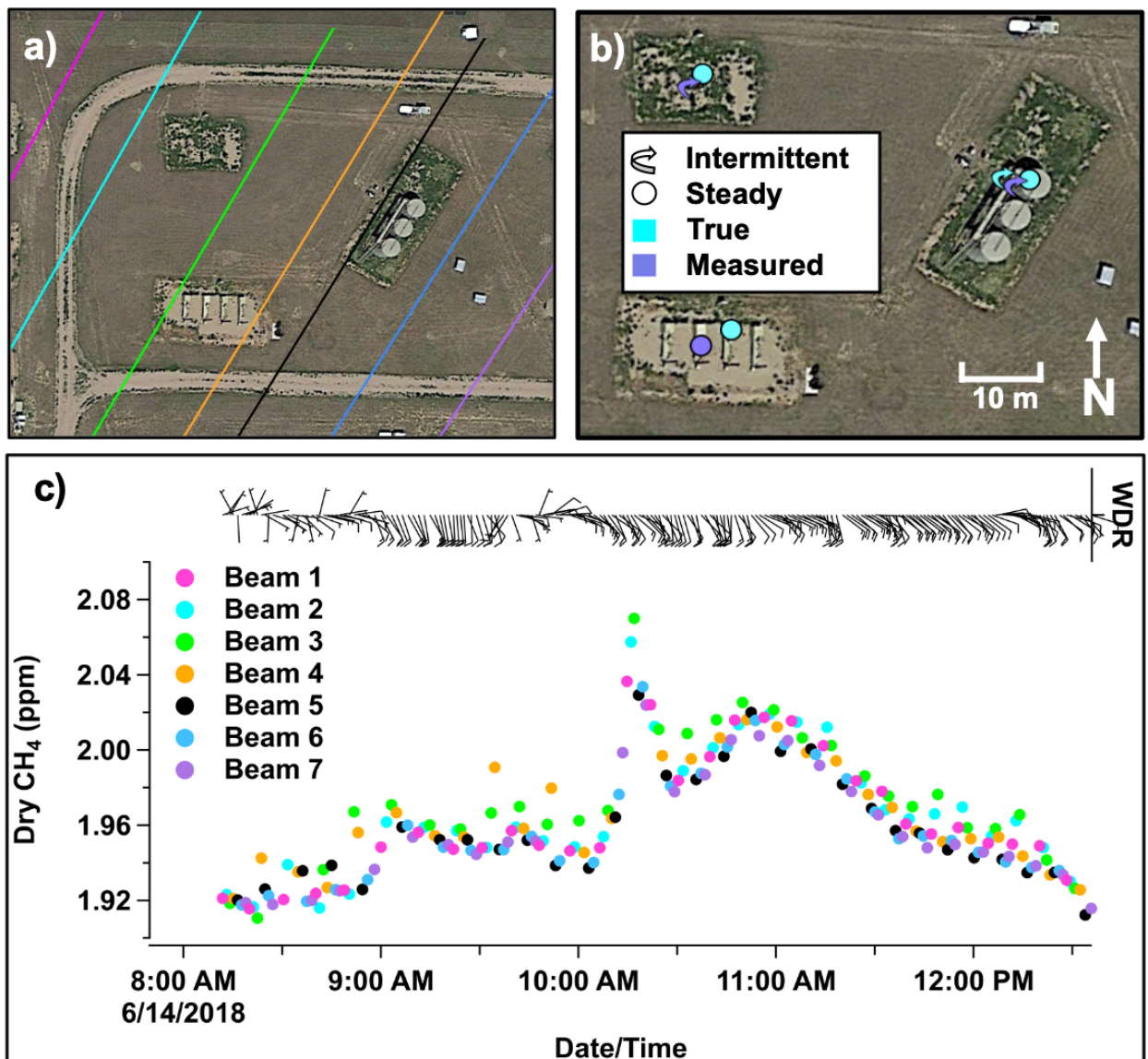

**Figure S1:** Specific results of Test #10. (a) Overview of the testing configuration and beam paths (colored lines) relative to equipment batteries. (b) Summary of measured (dark blue) and true (light blue) emissions. Emissions are further distinguished between steady (circles) and intermittent (curved arrows). (c) Time series of methane concentration measurements for the duration of the test. The beam path colors from panel (a) correspond with the point colors in panel (c). The slow variation of the overall concentration is due to site-wide time variation in the methane background, not instrument drift. Wind barbs: arrows point towards the direction from which the wind was blowing and the number of barbs indicates wind speed (1 barb = 2 m s$^{-1}$).

**Intermittent Emissions**

During tests at the METEC facility described here, the "C" difficulty tests utilized intermittent emission profiles (variable flow rate through time). These tests were designed to mimic observed emission profiles reported in the literature [2,3] that predominantly follow a repeating on/off pattern (as opposed to, for example, a single off/on/off emission event that is observed during liquid unloading). The METEC-engineered intermittency rates were "on" (flow rate ≠ 0) for a specific period of time, followed by "off" (flow rate = 0) for a specified period time, and the pattern was repeated for the duration of the test.

One of the "C" tests used exclusively intermittent emissions (whereas the other two "C" tests used both intermittent and steady emissions). The results from the exclusively intermittent test show the greatest difference between the true and measured emission rates (5.02 g min$^{-1}$ and 0.04 g min$^{-1}$, respectively). We perform a series of simulations and emission profile characterizations to evaluate why the system underestimated the emission rate for this particular test. To do this, we use the true emission rates and intermittency profiles (provided to us after submission and receipt of scored results by METEC) along with meteorological data collected at METEC in order to simulate the atmospheric concentration change in methane that would be measured by the DCS due to these known, true emissions over longer periods of time than was possible in the 4-hour testing window. We then process the simulated data using our standard inversion method, i.e., assuming that nothing is known regarding the number of leaks, location of leaks, or intermittency of leak rates, thereby introducing only those uncertainties inherent in our inversion framework related to detection, attribution, and rate variability and magnitude. Simulated fluctuations of ambient methane were not included in these synthetic data tests, in order to decouple this additional area of uncertainty. In this way, we isolate the sources of uncertainty and create a platform for testing whether our methodology introduces inherent biases due to intermittency. We perform inversions with this synthetic data using the meteorological measurements from the test itself (total of 4 hours), and then perform a series of additional tests that increase the amount of simulated monitoring time by adding 4-hours of additional actual meteorological measurements each time, until a total of 32 hours are tested.

This series of synthetic tests are generated at the battery level for each of the "C" difficulty tests and the results of the inversion for each test are summarized in figure S2. All emissions are presented as the error between the measure and true emission rate. We find that the inclusion of additional data, which simulates measuring for additional time, aids in the quantification of these intermittent emissions, and that under the simulated monitoring conditions (which exclude variation in the ambient methane concentrations), the true emission rate is reached within ~14-16 hours of total elapsed time.

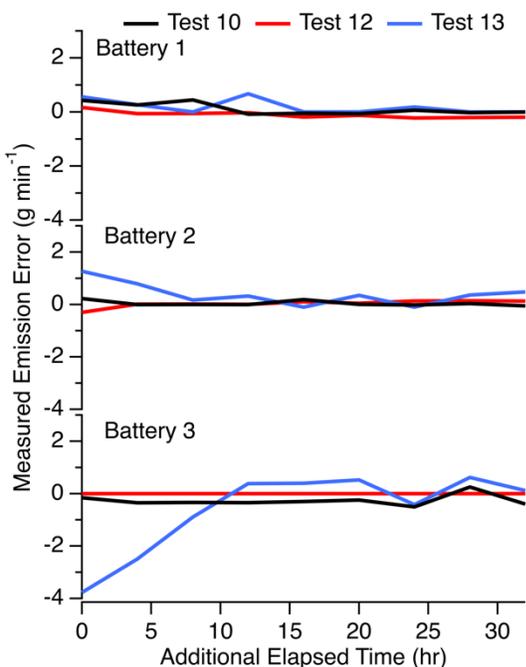

**Figure S2:** Results from the synthetic data tests investigating intermittent emissions showing the difference between the measured and true emissions as a function of time. Three different tests are simulated (all containing intermittent emissions) and are denoted by a Test ID (see figure 2) and distinguished by different colors for the traces. Each of the different batteries are also separated for each test.